\documentclass{article}
\usepackage{amsmath}
\usepackage{appendix}
\usepackage{amssymb}
\usepackage{adjustbox}
\usepackage{authblk}
\usepackage[style=numeric,maxnames=1,uniquelist=false,sorting=none]{biblatex}
\usepackage{graphicx} 

\title{Misaligned rings around minor planets with moons}
\author[1,2]{Barnabás Deme}

\affil[1]{Baja Astronomical Observatory of SZTE University, Szegedi út, Kt. 766,
Hungary, H-6500}
\affil[2]{ELTE E{\"o}tv{\"o}s Lor{\'a}nd University, Institute of Physics and Astronomy, Department of Astronomy, P{\'a}zm{\'a}ny P{\'e}ter stny. 1/A, H-1117, Budapest, Hungary}

\begin{document}

\maketitle

\begin{abstract}
Recent observations have confirmed the existence of rings around minor bodies in the outer Solar System. These objects may possess satellites as well. Here we investigate the interaction between such rings and satellites. We show that the perturbations from the moons may efficiently lead to off-equatorial rings around minor bodies like trans-neptunian objects or centaurs.  
\end{abstract}

\section{Introduction}

Rings around planetary bodies have intriguing dynamics. They are the outcome of small particles settling down to a thin plane due to kinetic energy losses while preserving their angular momentum \cite{Sicardy2006}. Their structure is shaped by internal collisions, interactions with solar radiation, resonances with moons, etc. \cite{Goldreich1982,murraydermott99}. It was long believed that they reside solely around gas giants. This paradigm changed with the discovery of ring systems around minor bodies in the outskirts of the solar system. There are currently four known minor planet systems which are ringed: two centaurs, Chariklo \cite{BragaRibas2014, Morgado2021} and Chiron \cite{Ortiz2015, Ruprecht2015} and two dwarf planets, Haumea \cite{Ortiz2017} and Quaoar \cite{Morgado2023}. 

The dynamics of minor planet rings differs from the giant planet counterparts in several aspects. For example, Chariklo and Haumea are non-axisymmetric, hence complicated resonant couplings arise between their spin and ring \cite{Sicardy2019}. Also, the ring of Quaoar lies outside its Roche radius, a feature that is yet to be explained \cite{Morgado2023}.

Many of the solar system's minor bodies, including those in the outer regions, possess moons \cite{Merline2002}. In several cases, the companion has a comparable size hence these systems are referred to as binary minor planets. They might have formed via gravitational capture \cite{Richardson2006} or in situ in the protoplanetary disk \cite{Nesvorny2010}. These satellites have fundamental significance in the dynamics of possible rings: they open gaps in it, sharpen their edges, stabilize them, etc. \cite{murraydermott99}.

The aim of this letter is to investigate another aspect of the interaction between moons and rings around minor bodies. On the one hand, these satellites/binary companions may contribute to the global gravitational field of the minor planet by a considerable amount; on the other hand, especially when they are captured, they can be significantly inclined with respect to the main body's equator. Consequently, they may lead to the formation of tilted, i.e. off-equatorial rings. Here we explore the parameter space where such misaligned rings may be present.

The structure of this paper is as follows. In Sec.~\ref{sec:ring_dyn} we outline the basic mathematical framework for studying ring dynamics and derive the equations for off-equatorial rings. In Sec.~\ref{sec:discussion} we discuss and visualize the results. App.~\ref{app:ai} provides some more details about the long-term precession of rings around minor bodies.

\section{Circular ring dynamics}\label{sec:ring_dyn}

For the sake of simplicity, we assume that rings are circular, i.e. the eccentricity of ring particles is zero. The vectorial equations of motion are then analogous to Euler's gyroscopic equations and read \cite{milankovitch39}
\begin{equation}\label{eq:eom}
    \dot{\mathbf{L}}=\boldsymbol{\Omega} \times \mathbf{L},
\end{equation}
where $\mathbf{L}$ is the angular momentum of a ring particle and $\boldsymbol{\Omega}$ is the precession vector, which is parallel with the axis of precession and its magnitude is the precession rate. Clearly, if $\mathbf{L}$ is parallel with $\boldsymbol{\Omega}$, i.e., if the particles orbit in the plane perpendicular to $\boldsymbol{\Omega}$, the angular momentum vector is constant. Due to the collisional dissipation of kinetic energy, ring particles settle down to this plane, called the Laplace surface \cite{Tremaine2009}. 

The precession vector is normally divided into two parts: $\boldsymbol{\Omega}=\boldsymbol{\Omega}_{J_2}+\boldsymbol{\Omega}_\mathrm{p}$, where
\begin{align}
    \boldsymbol{\Omega}_{J_2}&=-\frac{3}{2}\frac{\sqrt{\mathcal{G}m}J_2R^2}{a^{7/2}}(\mathbf{n}\cdot\hat{\mathbf{L}})\mathbf{n}, \label{eq:J2} \\
    \boldsymbol{\Omega}_\mathrm{p}&=-\frac{3}{4}\frac{\sqrt{\mathcal{G}}m_\mathrm{p} a^{3/2}}{\sqrt{m}a_\mathrm{p}^3(1-e_\mathrm{p}^2)^{3/2}}(\mathbf{n}_\mathrm{p} \cdot \hat{\mathbf{L}})\mathbf{n}_\mathrm{p}, \label{eq:p}
\end{align}
and $\hat{\mathbf{L}}$ is the normalized angular momentum, $\mathcal{G}$ is the gravitational constant, $m$ is the primary's (i.e. the ringed body's) mass, $J_2$ is its oblateness, $R$ is its characteristic radius, $\mathbf{n}$ is the unit vector along its rotational axis and $a$ is the ring's radius. The quantities $m_\mathrm{p}$, $a_\mathrm{p}$, $e_\mathrm{p}$ and $\mathbf{n}_\mathrm{p}$ are the perturbing object's mass, semi-major axis, eccentricity and orbital normal vector, respectively. $\boldsymbol{\Omega}_{J_2}$ and $\boldsymbol{\Omega}_\mathrm{p}$ account for nodal precession due to the primary's oblateness and the distant perturber's gravitational perturbation, respectively. Note that both effects are quadrupolar (i.e. the perturbation potentials scale with 1/distance$^3$) and secular (i.e. they do not depend on fast angles, like the mean anomaly). 

Another important feature of rings is that they must be located within the so-called Roche radius, where tidal effects prevent them from forming moons. Although depending on several complicated factors, like the densities of the planetary bodies, the thumb rule is that the Roche radius is on the order of the primary's radius; e.g., for equal densities it is $\sim1.44R$ \cite{murraydermott99}.

The inclination of the ring, i.e. the angle by which it is tilted from a reference plane, is the net result of the competing effects described by Eqs.~\eqref{eq:J2}-\eqref{eq:p}. In order to get a quantitative measure, we evaluate the precessions rates at the Roche radius ($a=1.44R$) and investigate their ratio
\begin{equation}\label{eq:ratio}
    \frac{\Omega_{J_2}}{\Omega_\mathrm{p}}\approx0.323\frac{m}{m_\mathrm{p}}\left(\frac{a_\mathrm{p}}{R}\right)^3 J_2 (1-e_\mathrm{p}^2)^{3/2}\frac{\mathbf{n}\cdot\hat{\mathbf{L}}}{\mathbf{n}_\mathrm{p}\cdot\hat{\mathbf{L}}},
\end{equation}
where $\Omega=|\boldsymbol{\Omega}|$. If this ratio is much smaller than 1, then the the distant perturber's effect dominates over the oblateness and thus dictates the inclination of the ring, making it possibly off-equatorial. Note that the Hill stability of the ring further requires \cite{Grishin2017}
\begin{equation}\label{eq:hill}
    \frac{m}{3m_\mathrm{p}} \left(\frac{a_\mathrm{p}(1-e_\mathrm{p})}{R}\right)^3 > 1.
\end{equation}

Now we seek for the condition for the ring to lie in the perturber's plane, i.e. $\mathbf{n}_\mathrm{p}\cdot\hat{\mathbf{L}}=1$. Demanding the ratio in Eq.~\eqref{eq:ratio} to be smaller than 1 along with the stability in Eq.~\eqref{eq:hill}, we obtain

\begin{equation}\label{eq:main}
    \frac{3}{(1-e_\mathrm{p})^3}<\frac{m}{m_\mathrm{p}}\left(\frac{a_\mathrm{p}}{R}\right)^3<\frac{3.1}{ J_2 (1-e_\mathrm{p}^2)^{3/2} \cos i},
\end{equation}
where $\cos i=\mathbf{n}\cdot\hat{\mathbf{L}}=\mathbf{n}\cdot\mathbf{n}_\mathrm{p}$.

As an illustrative example, let us investigate the case of Saturn, where the distant perturber is the Sun. Ignoring the factors with the eccentricity and inclination, which have a negligible effect, we get $(m/m_\mathrm{p})(a_\mathrm{p}/R)^3\sim10^9$ and $1/J_2\sim10^2$, hence Hill stability holds but the distant Sun's gravitational perturbation is completely quenched by Saturn's oblateness. As a conclusion, it is the latter that sets the plane of the rings \cite[see e.g. Fig. 5 in][]{Burns1979}.

\section{Discussion}\label{sec:discussion}

\begin{figure}[htbp!]
    \begin{center}
    \includegraphics[width=0.95\textwidth]{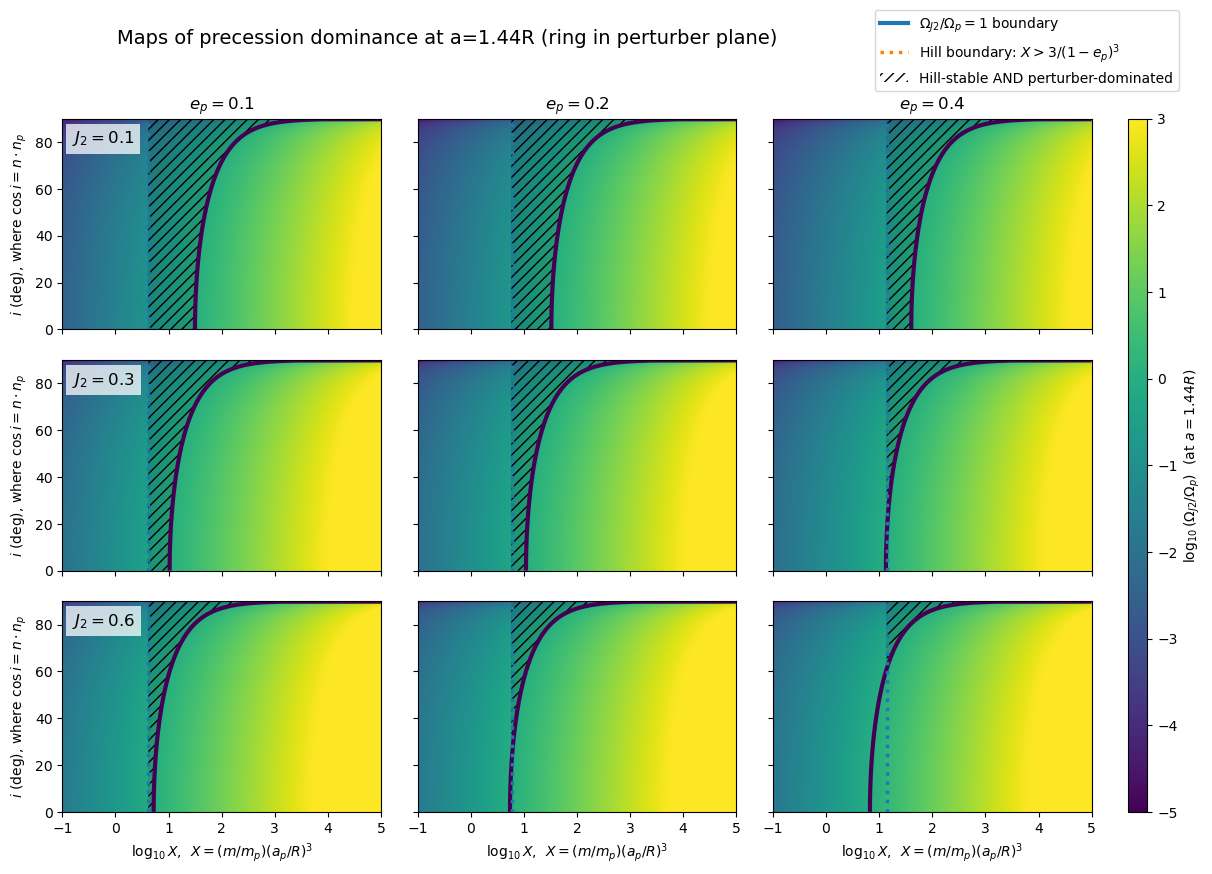}
    \caption{Heat map showing the relative strength of the primary's oblateness (yellow) and a satellite's perturbation (blue). The solid line separates the regions where any of the two effects become dominant. The dotted line marks the boundary of Hill stability. Rings in the shaded region are tilted as the result of a satellite's perturbation.}
    \label{fig:prec_dom}
    \end{center}
\end{figure}

Now we explore if rings around minor bodies can be tilted as a result of the perturbation from a satellite. Parameters which satisfy the criteria in Eq.~\eqref{eq:main} are in the shaded region of Fig.~\ref{fig:prec_dom}. Colors encode the relative strength of oblateness and satellite perturbations, while lines denote the thresholds of stability and satellite perturbation dominance.

The second inequality of Eq.~\eqref{eq:main} is trivially satisfied if the primary's oblateness vanishes, i.e. $J_2=0$. However, although it is a useful simplification in many studies \cite[see e.g.][]{Regaly2025}, it is obviously unphysical, especially for bodies which are highly non-spherical. For example, Ref.~\cite{Winter2023} assumes $J_2\approx0.13$ for Chariklo, the first minor body discovered to have a ring. Nevertheless, the region of parameters resulting in off-equatorial rings in Fig.~\ref{fig:prec_dom} becomes considerably wider as the oblateness decreases, as expected.

\begin{figure}[htbp!]
    \begin{center}
    \includegraphics[width=0.95\textwidth]{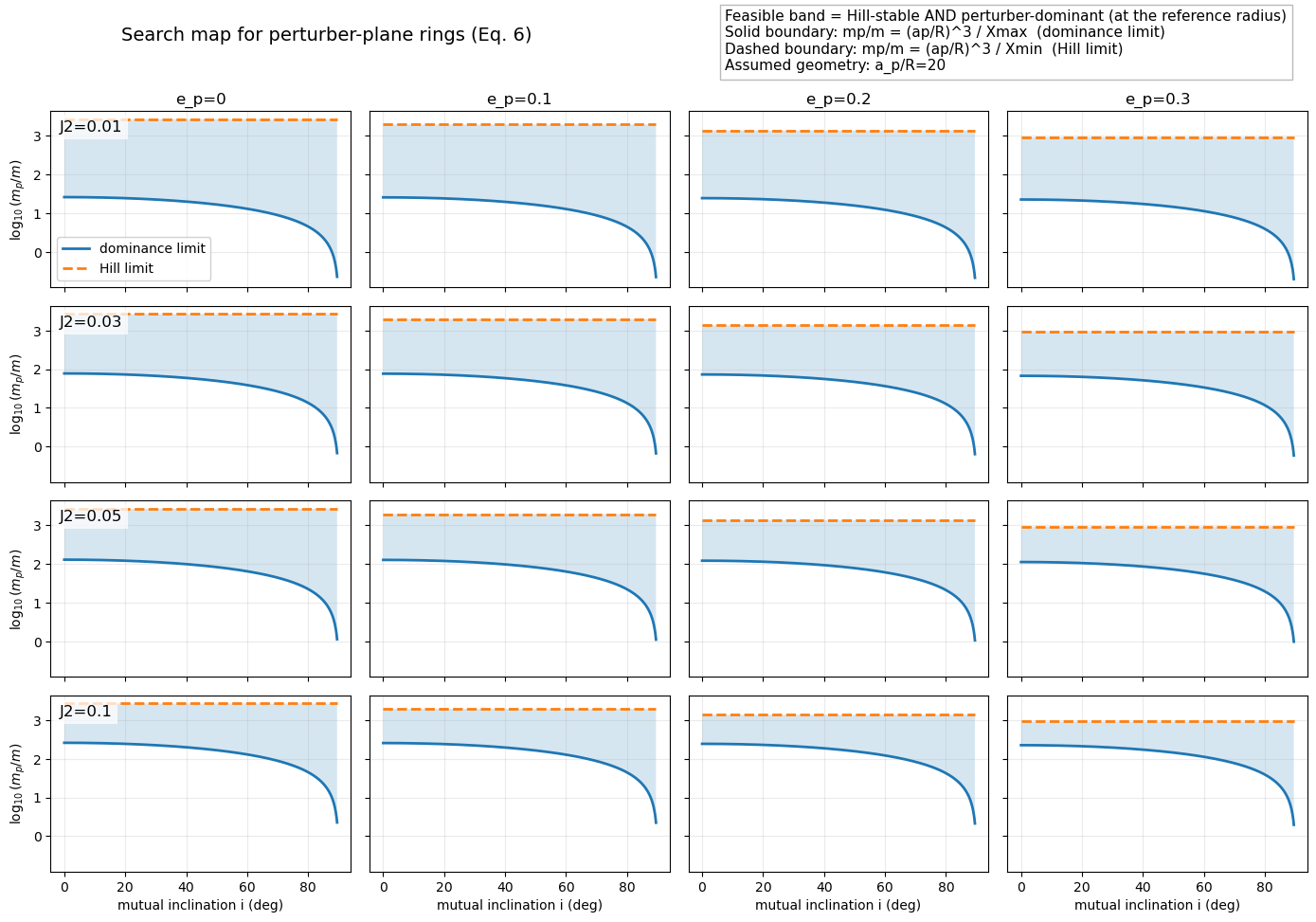}
    \caption{Allowed mass ratio and inclination parameters for off-equatorial rings by Eq.~\eqref{eq:main}. The solid and dashed lines have the same meaning as in Fig.~\ref{fig:prec_dom}. The available parameter space is clearly wider for higher inclinations.}
    \label{fig:search_map_inc}
    \end{center}
\end{figure}

Inclination can also be crucial. Importantly, even in the complete absence of external perturbers, planetary oblateness alone can result in polar, i.e. off-equatorial orbits, as is reflected by the $1/\cos i$ factor \cite{Dobr1989}. Indeed, $\mathbf{n}\cdot \mathbf{L}=0$ results in $\dot{\mathbf{L}}=0$, see Eqs.~\eqref{eq:eom}-\eqref{eq:J2}. This is reflected by the widening of the shaded region in the panels of Fig.~\ref{fig:prec_dom} as higher inclinations are reached. It is further demonstrated in Fig.~\ref{fig:search_map_inc}, where the mass ratio is plotted against the inclination of the satellite. Note that we made a conservative assumption in deriving Eq.~\eqref{eq:main}, where we set $\mathbf{n}_\mathrm{p}\cdot\hat{\mathbf{L}}=1$. It may happen that the ring inclinations is governed by a satellite yet its plane does not exactly fit to that of the perturber. It means that off-equatorial rings may be more common than what is hinted in this work. However, one should take into account that if a ring is sufficiently inclined with respect to a distant perturber and eccentric, then it may be subject to destabilization via Kozai oscillations \cite{Kozai1962, Naoz2016}.  

\begin{figure}[htbp!]
    \begin{center}
    \includegraphics[width=0.95\textwidth]{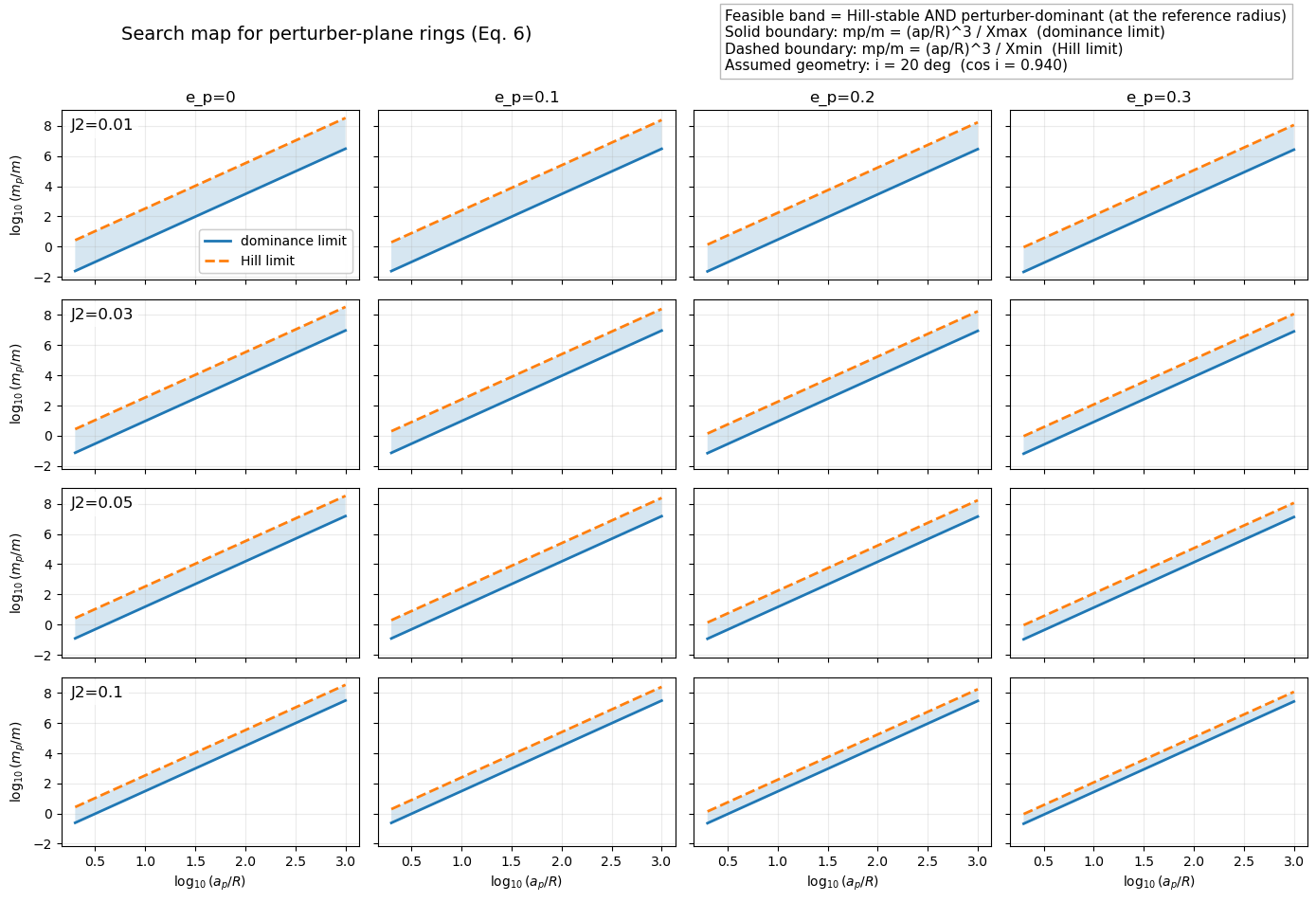}
    \caption{The same as Fig.~\ref{fig:search_map_inc} but for semi-major axis instead of eccentriciy. Binary minor bodies lie in the lower left, while ringed moons around gas giants in the upper right part of the blue-shaded stripe.}
    \label{fig:search_map}
    \end{center}
\end{figure}

In order to have an off-equatorial ring, i.e. satisfying Eq.~\eqref{eq:main}, one needs to decrease $m/m_\mathrm{p}$ or $a_\mathrm{p}/R$, although only to the extent that Hill stability holds. This latter requirement breaks down easily with increasing eccentricity, as can be seen in the right panels of Fig.~\ref{fig:prec_dom}. Also, for obvious reasons, the perturber's semi-major axis cannot decrease below the primary's radius. However, taking into account minor bodies in the solar system, there is no strict constraint for the masses and tilted rings can be possible in a variety of configurations, as shown in Fig.~\ref{fig:search_map}. For example, binary minor planets have $m\sim m_\mathrm{p}$. However, $m\ll m_\mathrm{p}$ is also possible, as in the case of rings around moons, given they are sufficiently far away from their host planet. Ref.~\cite{Suc2024} investigates this configuration, although assuming $J_2=0$.

Although the oblateness has negligible effect on off-equatorial rings by assumption, it may influence the dynamics of the perturber. In particular, the condition $\Omega_{J_2}(\mathrm{ring})\ll\Omega_{J_2}(\mathrm{perturber})$, i.e. when we assume that oblateness perturbation on the perturber is much larger than that on the ring, translates to
\begin{equation}
    \sqrt{m}R^{7/2}\ll\sqrt{m+m_\mathrm{p}}a_\mathrm{p}^{7/2}(1-e_\mathrm{p}^2)^2,
\end{equation}
where in the formula of $\boldsymbol{\Omega}_{J_2}(\mathrm{perturber})$ we made the substitution $m\to m+m_\mathrm{p}$. In this case, the orbital normal vector $\mathbf{n}_\mathrm{p}$ precesses with a rate that is related to the orbital frequency of the ring particles ($n=\sqrt{\mathcal{G}(m+m_\mathrm{p})a^3}$) as
\begin{equation}
    \frac{\Omega_{J_2}(\mathrm{perturber})}{n}\approx0.868 J_2\left(\frac{R}{a_\mathrm{p}}\right)^{7/2}\frac{\cos i}{(1-e_\mathrm{p}^2)^2},
\end{equation}
which is small (given the eccentricity is not too large due to Hill stability). Such a slow precession leaves the angle between $\boldsymbol{\Omega}$ and $\mathbf{L}$ (which is zero) adiabatically invariant, as shown heuristically in App.~\ref{app:ai}. Physically, it means that the slowly precessing perturber drags the tilted plane with itself.

In conclusion, minor bodies may host rings which are significantly misaligned with respect to their equator as a result of a perturbing satellite. More precisely, the Laplace surface may considerably deviate from the equatorial plane within the Roche radius. Given the increasing number of such discoveries in the near future thanks to LSST, considering such configurations may be important in accurately modeling the dynamics of rings around minor bodies.

\section*{Acknowledgments}
The author acknowledges the financial support of the Hungarian National
Research, Development and Innovation Office -- NKFIH Grant OTKA K-147131. He thanks E. Forgács-Dajka for her help in producing the figures. 

\printbibliography

\appendix

\section{The adiabatic invariance of the plane's tiltedness}\label{app:ai}

Here we investigate the long-term behavior of the solution of Eq.~\eqref{eq:eom}, if the precession vector is subject to slow changes. Without the loss of generality we assume that $\boldsymbol{\Omega}$ initially points in the $z$-direction. Then the solution of Eq.~\eqref{eq:eom} is
\begin{equation}\label{eq:L_solution}
    \mathbf{L}= \left(
        \begin{array}{c}
        L_\perp \cos(\Omega t+\phi_) \\
        L_\perp \sin(\Omega t+\phi_) \\
        L_\parallel \\
    \end{array}
    \right),               
\end{equation}
where $L_\perp$ and $L_\parallel$ are the perpendicular and parallel components of $\mathbf{L}$ and $\phi$ is the initial phase of precession. Here we heuristically show that a slowly changing $\boldsymbol{\Omega}$ drags the precessing $\mathbf{L}$ with itself.

Let us assume that the precession vector slowly changes as  $\boldsymbol{\Omega}=\boldsymbol{\Omega}_0+\epsilon \boldsymbol{\beta} t$ with $\boldsymbol{\beta}$ being constant. It follows to first order in $\epsilon$ that 
\begin{equation}\label{eq:Omega_sc}
    \Omega = \left[(\boldsymbol{\Omega}_0+\epsilon \boldsymbol{\beta} t)\cdot(\boldsymbol{\Omega}_0+\epsilon \boldsymbol{\beta} t)\right]^{1/2}\approx \Omega_0+\epsilon\frac{\boldsymbol{\Omega}_0\cdot\boldsymbol{\beta}t}{\Omega_0}=\Omega_0+\epsilon \beta_z t.
\end{equation}
The time derivative of $\boldsymbol{\Omega}\cdot \mathbf{L}$ is 
\begin{equation}
    \tfrac{\mathrm{d}}{\mathrm{d}t}(\boldsymbol{\Omega}\cdot \mathbf{L})=\dot{\boldsymbol{\Omega}}\cdot \mathbf{L}+\boldsymbol{\Omega}\cdot \dot{\mathbf{L}}=\epsilon\boldsymbol{\beta}\cdot \mathbf{L},
\end{equation}
where the second term in the second equality vanishes due to the equation of motion \eqref{eq:eom}. Note that it is a periodic function due to Eq.~\eqref{eq:L_solution}. Its average over one cycle is
\begin{equation}
    \langle \tfrac{\mathrm{d}}{\mathrm{d}t}(\boldsymbol{\Omega}\cdot \mathbf{L}) \rangle  = \epsilon \beta_z L_\parallel .
\end{equation}
Using this, the time evolution of $\boldsymbol{\Omega}\cdot \mathbf{L}$ may be written up to first order in $\epsilon$ as
\begin{equation}\label{eq:bold_OL}
    \boldsymbol{\Omega}\cdot\mathbf{L}=\Omega_0L_\parallel+\epsilon \beta_z L_\parallel t.
\end{equation}

Now let us compute the derivatve of $\Omega L$:
\begin{equation}
    \tfrac{\mathrm{d}}{\mathrm{d} t}(\Omega L)=\dot{\Omega} L + \Omega \dot{ L}=\epsilon \beta_z L,
\end{equation}
where we used Eq.~\eqref{eq:Omega_sc} and $\dot{L}=0$ again as a result of Eq.~\eqref{eq:eom}. The time evolution of $\Omega L$ is then
\begin{equation}\label{eq:plain_OL}
    \Omega L =\Omega_0 L + \epsilon \beta_z L t.
\end{equation}

The cosine of the angle between $\boldsymbol{\Omega}$ and $\mathbf{L}$ is $\cos \alpha=\boldsymbol{\Omega}\cdot\mathbf{L}/(\Omega L)$. We substitute Eqs.~\eqref{eq:bold_OL} and \eqref{eq:plain_OL}:
\begin{equation}
    \cos \alpha = \left(\Omega_0L_\parallel+\epsilon \beta_z L_\parallel t\right)\left(\Omega_0 L + \epsilon \beta_z L t\right)^{-1}\approx\frac{L_\parallel}{L},
\end{equation}
hence $\alpha$ is an adiabatic invariant up to first order. If $\alpha=0$ initially, i.e. $\mathbf{L}$ is parallel with $\boldsymbol{\Omega}$, then it remains to be so, thus $\boldsymbol{\Omega}$ slowly drags the orthogonal (Laplace) plane with itself.

Note that the calculation here is quite heuristic. Higher order invariance may be shown with a more sophisticated derivation. Also, since Eq.~\eqref{eq:eom} is the prototypical system for the so-called Nambu mechanics \cite{Nambu1973}, it would be very interesting to discuss the aforementioned invariance within that framework.

\end{document}